\documentclass[aps,prd, twocolumn,superscriptaddress,preprintnumbers,floatfix,nofootinbib,notitlepage]{revtex4-1}

\usepackage{graphicx}
\usepackage{amsmath,amsfonts,amssymb}
\usepackage[colorlinks,urlcolor=black,citecolor=blue,linkcolor=black]{hyperref}
\usepackage{color}
\usepackage{verbatim}

\def\lsim{\mathrel{\raise.3ex\hbox{$<$\kern-.75em\lower1ex\hbox{$\sim$}}}}
\def\gsim{\mathrel{\raise.3ex\hbox{$>$\kern-.75em\lower1ex\hbox{$\sim$}}}}

\newcommand{\ie}{{\it i.e.}}
\newcommand{\eg}{{\it e.g.\,}}

\newcommand{\be}{\begin{equation}}
\newcommand{\ee}{\end{equation}}
\newcommand{\bea}{\begin{equation}\begin{aligned}}
\newcommand{\eea}{\end{aligned}\end{equation}}

\begin{document}

\title{Reply to "Comment on: Cosmological black holes are not described by \\ the Thakurta metric"
}

\author{Gert H\"utsi}
\email{gert.hutsi@to.ee}
\affiliation{Laboratory of High Energy and Computational Physics, NICPB, R\"avala pst. 10, 10143 Tallinn, Estonia}

\author{Tomi Koivisto}
\email{timoko@kth.se}
\affiliation{Laboratory of High Energy and Computational Physics, NICPB, R\"avala pst. 10, 10143 Tallinn, Estonia}
\affiliation{Laboratory of Theoretical Physics, University of Tartu, W. Ostwaldi 1, 50411 Tartu, Estonia}

\author{Martti Raidal}
\email{martti.raidal@cern.ch}
\affiliation{Laboratory of High Energy and Computational Physics, NICPB, R\"avala pst. 10, 10143 Tallinn, Estonia}

\author{Ville Vaskonen}
\email{vvaskonen@ifae.es}
\affiliation{Institut de Fisica d'Altes Energies (IFAE), The Barcelona Institute of Science and Technology, Campus UAB, 08193 Bellaterra (Barcelona), Spain}

\author{Hardi Veerm\"ae}
\email{hardi.veermae@cern.ch}
\affiliation{Laboratory of High Energy and Computational Physics, NICPB, R\"avala pst. 10, 10143 Tallinn, Estonia}

\begin{abstract}
In this reply, we address the comment~\cite{Boehm:2021kzq} to our recent paper~\cite{Hutsi:2021nvs}, where we argued that the Thakurta metric does not describe cosmological black holes. We clarify that the mass growth of Thakurta black holes is due to an influx of energy (\ie \, accretion), which, by definition, is not a feature of geometry. The conclusions of Ref.~\cite{Hutsi:2021nvs} are independent of the interpretation of this energy flux. We show that the average energy density of primordial Thakurta black holes scales as $a^{-2}$ and requires an unrealistic and fine-tuned energy transfer from a smooth dark matter component to the primordial black hole sector.
\end{abstract}

\maketitle

Recently, Boehm {\it et al.}~\cite{Boehm:2020jwd} claimed that cosmological black holes (BHs) are described by Thakurta metric and, therefore, for example, the LIGO-Virgo bounds on the possible primordial black hole (PBH) abundance (see~\cite{Hutsi:2020sol} for the latest constraint) would be invalid. In~\cite{Hutsi:2021nvs} we argued that the physical conditions which induce the Thakurta metric are not realized in any realistic cosmological scenario. Our criticism was questioned by Boehm {\it et al.} in~\cite{Boehm:2021kzq}, who believe that in~\cite{Hutsi:2021nvs} we misinterpreted the Thakurta metric. In this reply, we address the comments presented in Ref.~\cite{Boehm:2021kzq}.

The Misner-Sharp mass enclosed in a sphere of areal radius $R$ of a BH described by the Thakurta metric,
\be\label{M_MS}
    M_{\rm MS} 
    = ma(t) + \frac{4\pi}{3} \rho_{M}(R,t) R^3\,,
\ee
where $a$ denotes the scale factor and $\rho_{M}$ the energy density at areal radius $R>0$, consists of 2 components. The first component, $M \equiv ma,$ is present for arbitrarily small $R$ and thus describes a point-mass like contribution. The second component, where $\rho_{M}(R,t)$ is related to the Hubble parameter $H$ via $\rho_{M}(R,t) = 3H^2/(8\pi (1-2M/R))$, can be interpreted as the contribution of surrounding matter.

In~\cite{Hutsi:2021nvs} we argued that the Thakurta metric describes BHs accreting with an unphysically high rate. The comment~\cite{Boehm:2021kzq} claimed that the Thakurta metric does not imply fluid accretion but, instead, the mass growth is a feature of the spacetime geometry. Here we show that this claim does not hold up to scrutiny. 

In general, the 4-velocity of the cosmological fluid can be parametrized as ${\bf u}= \cosh\chi {\bf e}_0 + \sinh\chi{\bf e}_r$, and the consistency of the Thakurta geometry requires to take also into account the imperfect heat flow 4-vector  ${\bf q} = q(\sinh\chi{\bf e}_0 +  \cosh\chi {\bf e}_r)$, which is orthogonal to ${\bf u}$ and is responsible for the mass growth, \ie,~$\dot{M} \sim q$. As explained in Section IV.C of Ref.~\cite{Carrera:2009ve}, there are two contributions to $q=J_M+J_H$, the cosmological matter $J_M$ and the heat flow $J_H$. If the rapidity is exactly vanishing, $\chi=0$, the mass growth is formally solely due to the heat flow {\it into the BH,} and $J_M=0$. However, even an infinitesimal $\chi$ changes the solution {\it qualitatively} and allows for a physical interpretation. That is, any matter flow into the Thakurta BH invalidates the discontinuous, infinitely fine-tuned $\chi=0$ solution. Some inflow of matter is guaranteed in any realistic cosmology with matter such as baryons. In the small $\chi$ case, one finds~\cite{Carrera:2009ve}
\be
    J_M/J_H = -\frac{2\cosh^2\chi}{1+2\sinh^2\chi} \approx -2\,.
\ee
Thus, {\it the mass growth is due to the infalling cosmological matter}, which has to have double the rate to compensate for the {\it outflowing heat} required for the consistency of the solution.

Nevertheless, for the arguments of Ref.~\cite{Hutsi:2021nvs} to hold, it is sufficient that the Thakurta metric is sourced by a stress-energy tensor that contains a radial energy flux into the BH, which supports its mass growth with the accretion rate\footnote{By accretion we mean $\dot M > 0$. In general relativity, this implies an influx of energy (as opposed to a spontaneous generation of mass as in, \eg, the steady-state universe).}
\be\label{dotM}
    \dot{M} = H M.
\ee
We must stress that this energy flow is, by construction, a property of the source, not of geometry (even for $\chi=0$). The crucial point here is that the mass of the BH grows due to the redistribution of the energy density carried by the matter surrounding the BH. Whether such an energy flux is realized is ultimately determined by the properties of dark matter. No dark matter candidate is known to support this specific energy flow (especially if it is a pure heat flow). For example, in~\cite{Hutsi:2021nvs} we argue that if all dark matter consists of PBHs, then the surrounding matter making up $\rho_{M}$ is not described by a smooth stress-energy tensor and is not capable of causing such an energy flux. The issues with the scenario where all dark matter consists of PBHs were not addressed in the comment~\cite{Boehm:2021kzq}. 

In the comment~\cite{Boehm:2021kzq} the parameter $m$ was interpreted as a physical mass that contributes to global quantities such as the average energy density of PBHs, \ie, $\rho_{\rm PBH} = m\, n_{\rm PBH}$, where $n_{\rm PBH}$ the PBH number density. However, $m$ enters as a parameter of the Thakurta metric and does not represent the energy carried by the BH at times other than $a=1$. For example, one can consider the $H \to 0$ limit with an almost constant scale factor, in which case the BH mass is clearly $M = ma$. 

Moreover, since the stress-energy tensor sourcing an individual BH is known, it is possible to estimate the energy density in a universe filled with Thakurta BHs. To address constraints on PBHs, we are interested in times not much earlier than matter-radiation equality. Thus we can assume that the cosmological and event horizons are well separated, and any Hubble patch contains many PBHs whose separation is much larger than the size of the BH horizon. One way to describe such a universe is by gluing together several Thakurta spacetimes.\footnote{Gluing together only Thakurta patches clashes with energy conservation at their boundary because each patch requires an energy influx. Therefore a consistent description of a Hubble patch containing several Thakurta BHs is non-trivial and needs additional physical assumptions.} For our current purposes, it is sufficient to focus on the energy contained in each patch. This energy is given by the Misner-Sharp mass \eqref{M_MS} and, as discussed above, can be split into the mass of the BH $M$ and to a smooth component with energy density $\rho_{M,0} \approx  \lim_{R\to \infty} \rho_{M}(R)$. We stress that the energy density surrounding the Thakurta BH is due to a smooth matter component, \ie, not due to other BHs, so $\rho_{M,0}$ should not be identified with the PBH energy density. Therefore, a universe filled with Thakurta BHs would have an average energy density
\be
    \rho \approx \rho_{M,0} + m a \, n_{\rm PBH},
\ee
in contradiction with the statement made in the comment~\cite{Boehm:2021kzq}. According to the Thakurta metric, energy is transferred between the two components. Conservation of the number of PBHs, $n_{\rm PBH} \propto a^{-3}$, infers that the energy density of the PBH component is diluted as $a^{-2}$. Unless this scaling is due to a suitable transfer of energy between the PBHs and the smooth dark matter component so that their combined energy density scales as $a^{-3}$, Thakurta PBHs could not behave as dark matter.

To conclude, in this reply we have clarified that the Thakurta metric describes accreting BHs whose mass grows $\propto a$ and that must be accompanied by a smooth dark matter component which the BHs can accrete. Based on the arguments presented in~\cite{Hutsi:2021nvs}, the required accretion rate is not realistic and, therefore, the Thakurta metric does not describe cosmological black holes. We did also show that the energy density of Thakurta PBHs is $m a \, n_{\rm PBH}$ instead of $m n_{\rm PBH}$ as claimed in~\cite{Boehm:2021kzq}.


\vskip 0.5cm
\noindent
\emph{Acknowledgments.} This work was supported by the Estonian Research Council grants PRG356, PRG803, MOBTB135,  MOBJD381, MOBTT86 and MOBTT5, and by the EU through the European Regional Development Fund CoE program TK133 ``The Dark Side of the Universe." This work was also supported by the grants FPA2017-88915-P and SEV-2016-0588. IFAE is partially funded by the CERCA program of the Generalitat de Catalunya.

\appendix

\bibliography{PBH}

\begin{thebibliography}{5}%
\makeatletter
\providecommand \@ifxundefined [1]{%
 \@ifx{#1\undefined}
}%
\providecommand \@ifnum [1]{%
 \ifnum #1\expandafter \@firstoftwo
 \else \expandafter \@secondoftwo
 \fi
}%
\providecommand \@ifx [1]{%
 \ifx #1\expandafter \@firstoftwo
 \else \expandafter \@secondoftwo
 \fi
}%
\providecommand \natexlab [1]{#1}%
\providecommand \enquote  [1]{``#1''}%
\providecommand \bibnamefont  [1]{#1}%
\providecommand \bibfnamefont [1]{#1}%
\providecommand \citenamefont [1]{#1}%
\providecommand \href@noop [0]{\@secondoftwo}%
\providecommand \href [0]{\begingroup \@sanitize@url \@href}%
\providecommand \@href[1]{\@@startlink{#1}\@@href}%
\providecommand \@@href[1]{\endgroup#1\@@endlink}%
\providecommand \@sanitize@url [0]{\catcode `\\12\catcode `\$12\catcode
  `\&12\catcode `\#12\catcode `\^12\catcode `\_12\catcode `\%12\relax}%
\providecommand \@@startlink[1]{}%
\providecommand \@@endlink[0]{}%
\providecommand \url  [0]{\begingroup\@sanitize@url \@url }%
\providecommand \@url [1]{\endgroup\@href {#1}{\urlprefix }}%
\providecommand \urlprefix  [0]{URL }%
\providecommand \Eprint [0]{\href }%
\providecommand \doibase [0]{http://dx.doi.org/}%
\providecommand \selectlanguage [0]{\@gobble}%
\providecommand \bibinfo  [0]{\@secondoftwo}%
\providecommand \bibfield  [0]{\@secondoftwo}%
\providecommand \translation [1]{[#1]}%
\providecommand \BibitemOpen [0]{}%
\providecommand \bibitemStop [0]{}%
\providecommand \bibitemNoStop [0]{.\EOS\space}%
\providecommand \EOS [0]{\spacefactor3000\relax}%
\providecommand \BibitemShut  [1]{\csname bibitem#1\endcsname}%
\let\auto@bib@innerbib\@empty
\bibitem [{\citenamefont {Boehm}\ \emph
  {et~al.}(2021{\natexlab{a}})\citenamefont {Boehm}, \citenamefont
  {Kobakhidze}, \citenamefont {O'Hare}, \citenamefont {Picker},\ and\
  \citenamefont {Sakellariadou}}]{Boehm:2021kzq}%
  \BibitemOpen
  \bibfield  {author} {\bibinfo {author} {\bibfnamefont {C.}~\bibnamefont
  {Boehm}}, \bibinfo {author} {\bibfnamefont {A.}~\bibnamefont {Kobakhidze}},
  \bibinfo {author} {\bibfnamefont {C.~A.~J.}\ \bibnamefont {O'Hare}}, \bibinfo
  {author} {\bibfnamefont {Z.~S.~C.}\ \bibnamefont {Picker}}, \ and\ \bibinfo
  {author} {\bibfnamefont {M.}~\bibnamefont {Sakellariadou}},\ }\href@noop {}
  {\  (\bibinfo {year} {2021}{\natexlab{a}})},\ \Eprint
  {http://arxiv.org/abs/2105.14908} {arXiv:2105.14908 [astro-ph.CO]}
  \BibitemShut {NoStop}%
\bibitem [{\citenamefont {H\"utsi}\ \emph
  {et~al.}(2021{\natexlab{a}})\citenamefont {H\"utsi}, \citenamefont
  {Koivisto}, \citenamefont {Raidal}, \citenamefont {Vaskonen},\ and\
  \citenamefont {Veerm\"ae}}]{Hutsi:2021nvs}%
  \BibitemOpen
  \bibfield  {author} {\bibinfo {author} {\bibfnamefont {G.}~\bibnamefont
  {H\"utsi}}, \bibinfo {author} {\bibfnamefont {T.}~\bibnamefont {Koivisto}},
  \bibinfo {author} {\bibfnamefont {M.}~\bibnamefont {Raidal}}, \bibinfo
  {author} {\bibfnamefont {V.}~\bibnamefont {Vaskonen}}, \ and\ \bibinfo
  {author} {\bibfnamefont {H.}~\bibnamefont {Veerm\"ae}},\ }\href@noop {} {\
  (\bibinfo {year} {2021}{\natexlab{a}})},\ \Eprint
  {http://arxiv.org/abs/2105.09328} {arXiv:2105.09328 [astro-ph.CO]}
  \BibitemShut {NoStop}%
\bibitem [{\citenamefont {Boehm}\ \emph
  {et~al.}(2021{\natexlab{b}})\citenamefont {Boehm}, \citenamefont
  {Kobakhidze}, \citenamefont {O'hare}, \citenamefont {Picker},\ and\
  \citenamefont {Sakellariadou}}]{Boehm:2020jwd}%
  \BibitemOpen
  \bibfield  {author} {\bibinfo {author} {\bibfnamefont {C.}~\bibnamefont
  {Boehm}}, \bibinfo {author} {\bibfnamefont {A.}~\bibnamefont {Kobakhidze}},
  \bibinfo {author} {\bibfnamefont {C.~A.~J.}\ \bibnamefont {O'hare}}, \bibinfo
  {author} {\bibfnamefont {Z.~S.~C.}\ \bibnamefont {Picker}}, \ and\ \bibinfo
  {author} {\bibfnamefont {M.}~\bibnamefont {Sakellariadou}},\ }\href {\doibase
  10.1088/1475-7516/2021/03/078} {\bibfield  {journal} {\bibinfo  {journal}
  {JCAP}\ }\textbf {\bibinfo {volume} {03}},\ \bibinfo {pages} {078} (\bibinfo
  {year} {2021}{\natexlab{b}})},\ \Eprint {http://arxiv.org/abs/2008.10743}
  {arXiv:2008.10743 [astro-ph.CO]} \BibitemShut {NoStop}%
\bibitem [{\citenamefont {H\"utsi}\ \emph
  {et~al.}(2021{\natexlab{b}})\citenamefont {H\"utsi}, \citenamefont {Raidal},
  \citenamefont {Vaskonen},\ and\ \citenamefont {Veerm\"ae}}]{Hutsi:2020sol}%
  \BibitemOpen
  \bibfield  {author} {\bibinfo {author} {\bibfnamefont {G.}~\bibnamefont
  {H\"utsi}}, \bibinfo {author} {\bibfnamefont {M.}~\bibnamefont {Raidal}},
  \bibinfo {author} {\bibfnamefont {V.}~\bibnamefont {Vaskonen}}, \ and\
  \bibinfo {author} {\bibfnamefont {H.}~\bibnamefont {Veerm\"ae}},\ }\href
  {\doibase 10.1088/1475-7516/2021/03/068} {\bibfield  {journal} {\bibinfo
  {journal} {JCAP}\ }\textbf {\bibinfo {volume} {03}},\ \bibinfo {pages} {068}
  (\bibinfo {year} {2021}{\natexlab{b}})},\ \Eprint
  {http://arxiv.org/abs/2012.02786} {arXiv:2012.02786 [astro-ph.CO]}
  \BibitemShut {NoStop}%
\bibitem [{\citenamefont {Carrera}\ and\ \citenamefont
  {Giulini}(2010)}]{Carrera:2009ve}%
  \BibitemOpen
  \bibfield  {author} {\bibinfo {author} {\bibfnamefont {M.}~\bibnamefont
  {Carrera}}\ and\ \bibinfo {author} {\bibfnamefont {D.}~\bibnamefont
  {Giulini}},\ }\href {\doibase 10.1103/PhysRevD.81.043521} {\bibfield
  {journal} {\bibinfo  {journal} {Phys. Rev. D}\ }\textbf {\bibinfo {volume}
  {81}},\ \bibinfo {pages} {043521} (\bibinfo {year} {2010})},\ \Eprint
  {http://arxiv.org/abs/0908.3101} {arXiv:0908.3101 [gr-qc]} \BibitemShut
  {NoStop}%
\end{thebibliography}%
\end{document}